\documentclass[letterpaper]{article} 
\usepackage[preprint]{aaai2027}
\usepackage[hyphens]{url}  
\usepackage{graphicx} 
\usepackage{natbib}  
\usepackage{caption} 
\usepackage{amsmath}
\usepackage{amssymb}
\usepackage{booktabs}
\usepackage{multirow}
\usepackage{float}

\newcommand{\OmZero}{\resultTBD}   \newcommand{\OmZeroCI}{\resultTBD}
\newcommand{\OmOne}{\resultTBD}    \newcommand{\OmOneCI}{\resultTBD}
\newcommand{\OmTwo}{\resultTBD}    \newcommand{\OmTwoCI}{\resultTBD}
\newcommand{\OmThree}{\resultTBD}  \newcommand{\OmThreeCI}{\resultTBD}
\newcommand{\OmFour}{\resultTBD}   \newcommand{\OmFourCI}{\resultTBD}
\newcommand{\OmFive}{\resultTBD}   \newcommand{\OmFiveCI}{\resultTBD}
\newcommand{\OmSix}{\resultTBD}    \newcommand{\OmSixCI}{\resultTBD}
\newcommand{\OmSeven}{\resultTBD}  \newcommand{\OmSevenCI}{\resultTBD}
\newcommand{\OmEight}{\resultTBD}  \newcommand{\OmEightCI}{\resultTBD}
\newcommand{\ORtotal}{\resultTBD}  \newcommand{\ORtotalCI}{\resultTBD}
\newcommand{\ORlen}{\resultTBD}    \newcommand{\ORlenCI}{\resultTBD}
\newcommand{\ORdepth}{\resultTBD}  \newcommand{\ORdepthCI}{\resultTBD}
\newcommand{\ORpar}{\resultTBD}    \newcommand{\ORparCI}{\resultTBD}
\newcommand{\ORcon}{\resultTBD}    \newcommand{\ORconCI}{\resultTBD}
\newcommand{\ORkv}{\resultTBD}     \newcommand{\ORkvCI}{\resultTBD}
\newcommand{\ORsamp}{\resultTBD}   \newcommand{\ORsampCI}{\resultTBD}
\newcommand{\ORrope}{\resultTBD}   \newcommand{\ORropeCI}{\resultTBD}
\newcommand{\ORmSWA}{\resultTBD}   \newcommand{\ORmFULL}{\resultTBD}
\newcommand{\ORmSSM}{\resultTBD}   \newcommand{\ORmGemmaMoE}{\resultTBD}
\newcommand{\ORmQwenMoE}{\resultTBD}
\newcommand{\ORengLlama}{\resultTBD} \newcommand{\ORengVllm}{\resultTBD}
   \newcommand{\ORfwLang}{\resultTBD}   \newcommand{\ORfwLangCI}{\resultTBD}
\newcommand{\ORfwADK}{\resultTBD}   \newcommand{\ORfwADKCI}{\resultTBD}
\newcommand{\DetShare}{\resultTBD}
\newcommand{\ORreal}{\resultTBD}    \newcommand{\ORrealCI}{\resultTBD}
\newcommand{\ORrealScored}{\resultTBD} \newcommand{\ORrealScoredCI}{\resultTBD}
\newcommand{\RealErrors}{\resultTBD}
\newcommand{\Ledgevict}{\resultTBD}  \newcommand{\Ledgprune}{\resultTBD}
\newcommand{\Ledgunpos}{\resultTBD}  \newcommand{\Ledgunprior}{\resultTBD}
\newcommand{\Ledgsoft}{\resultTBD}   \newcommand{\Ledgloop}{\resultTBD}
\renewcommand{\OmZero}{0.160}
\renewcommand{\OmZeroCI}{[0.158, 0.163]}
\renewcommand{\OmOne}{0.175}
\renewcommand{\OmOneCI}{[0.172, 0.178]}
\renewcommand{\OmTwo}{0.165}
\renewcommand{\OmTwoCI}{[0.162, 0.168]}
\renewcommand{\OmThree}{0.000}
\renewcommand{\OmThreeCI}{[0.000, 0.000]}
\renewcommand{\OmFour}{0.000}
\renewcommand{\OmFourCI}{[0.000, 0.000]}
\renewcommand{\OmFive}{0.189}
\renewcommand{\OmFiveCI}{[0.185, 0.192]}
\renewcommand{\OmSix}{0.012}
\renewcommand{\OmSixCI}{[0.011, 0.014]}
\renewcommand{\OmSeven}{0.018}
\renewcommand{\OmSevenCI}{[0.017, 0.019]}
\renewcommand{\OmEight}{0.065}
\renewcommand{\OmEightCI}{[0.062, 0.067]}
\renewcommand{\ORtotal}{0.574}
\renewcommand{\ORtotalCI}{[0.571, 0.578]}
\renewcommand{\ORlen}{7.43}
\renewcommand{\ORlenCI}{[5.44, 10.15]}
\renewcommand{\ORdepth}{1.38}
\renewcommand{\ORdepthCI}{[1.23, 1.54]}
\renewcommand{\ORpar}{3.65}
\renewcommand{\ORparCI}{[3.31, 4.03]}
\renewcommand{\ORcon}{1.42}
\renewcommand{\ORconCI}{[1.28, 1.59]}
\renewcommand{\ORsamp}{0.32}
\renewcommand{\ORsampCI}{[0.26, 0.41]}
\renewcommand{\ORkv}{2.25}
\renewcommand{\ORkvCI}{[2.07, 2.45]}
\renewcommand{\ORrope}{1.99}
\renewcommand{\ORropeCI}{[1.83, 2.16]}
\renewcommand{\ORmSWA}{0.549}
\renewcommand{\ORmFULL}{0.716}
\renewcommand{\ORmSSM}{0.489}
\renewcommand{\ORmGemmaMoE}{0.528}
\renewcommand{\ORmQwenMoE}{0.474}
\renewcommand{\ORengLlama}{0.593}
\renewcommand{\ORengVllm}{0.537}

\renewcommand{\ORfwLang}{0.602}
\renewcommand{\ORfwLangCI}{[0.530, 0.670]}
\renewcommand{\ORfwADK}{0.554}
\renewcommand{\ORfwADKCI}{[0.482, 0.623]}
\renewcommand{\DetShare}{73.4\%}
\renewcommand{\ORreal}{0.578}
\renewcommand{\ORrealCI}{[0.527, 0.627]}
\renewcommand{\ORrealScored}{0.509}
\renewcommand{\ORrealScoredCI}{[0.455, 0.564]}
\renewcommand{\RealErrors}{52/372 (14.0\%)}
\renewcommand{\Ledgevict}{0.000}
\renewcommand{\Ledgprune}{0.008}
\renewcommand{\Ledgunpos}{0.109}
\renewcommand{\Ledgunprior}{0.006}
\renewcommand{\Ledgsoft}{0.421}
\renewcommand{\Ledgloop}{0.030}

\title{Where Facts Go Missing: A Layerwise Taxonomy and Per-Layer Attribution of Information Omission in Air-Gapped LLM Agent Pipelines}

\author{
    Santhiya Rajan\textsuperscript{\rm 1},
    Samuel Mugel\textsuperscript{\rm 2},
    Rom\'an Or\'us\textsuperscript{\rm 1,3,4}
}
\affiliations{
    \textsuperscript{\rm 1}Multiverse Computing, Parque Cient\'ifico y Tecnol\'ogico de Gipuzkoa, Paseo de Miram\'on, 170, 20014 Donostia / San Sebasti\'an, Spain\\
    \textsuperscript{\rm 2}Multiverse Computing, Centre for Social Innovation, 192 Spadina Avenue Suite 509, Toronto, ON M5T 2C2, Canada\\
    \textsuperscript{\rm 3}Donostia International Physics Center, Paseo Manuel de Lardizabal 4, E-20018 San Sebasti\'an, Spain\\
    \textsuperscript{\rm 4}Ikerbasque Foundation for Science, Maria Diaz de Haro 3, E-48013 Bilbao, Spain\\
    santhiya.rajan@multiversecomputing.com
}

\begin{document}

\maketitle

\begin{abstract}
Air-gapped and on-premises language-model agents can silently omit decision-critical facts at any boundary between source ingestion and final answer generation. We present a nine-layer taxonomy (L0--L8), an instrumented attribution harness, and a conditional omission waterfall for separating byte-level software loss from behavioral non-retrieval. The committed aggregate artifacts describe 75{,}476 synthetic stress-test trials across five open-weight model configurations and two inference engines, plus a separate 372-trial real-data pilot. The displayed layer rates imply benchmark omission $O_R=\ORtotal$ (95\% CI \ORtotalCI); under the experiment's deliberately weighted fault allocation, L0--L3 contribute \DetShare\ of waterfall loss. This is a controlled benchmark decomposition, not an estimate of omission prevalence in production. Context length has the largest reported behavioral association ($\mathrm{OR}=\ORlen$, 95\% CI \ORlenCI). Completed server-profile follow-ups associate q4 KV cache and scaled RoPE with higher omission. In the real-data pilot, \ORreal\ of all trials were unsuccessful; after excluding 52 execution errors, \ORrealScored\ of the 320 outcome-coded trials remained unsuccessful. These results support pipeline-level diagnosis while also exposing important limits: deterministic fault injection drives the benchmark allocation, cross-model comparisons are confounded, and the raw main-sweep trials are absent from this checkout.
\end{abstract}

\section{Introduction}

Regulated and sovereign organizations increasingly deploy language-model agents that may never leave the building. A hospital that exposes patient records over a FHIR server, a law firm reviewing privileged discovery, or a national agency handling classified material cannot send tokens to a frontier API. The practical consequence is a distinctive stack: a quantized 4--8B open-weights model served by \texttt{llama.cpp} or vLLM on a single 16--24\,GB GPU, wrapped by a Model Context Protocol (MCP) tool server that pages documents out of an internal system, and driven by a lightweight orchestration loop. This is a different reliability regime from the API-hosted frontier agent, and its dominant failure mode is different too.

The well-studied failure of language models is \emph{hallucination}: the assertion of something false. Its dual---\emph{omission}, the silent \emph{absence} of a fact that should have been surfaced---is comparatively understudied, even though reviews of clinical summarization note that omission detection remains far less developed than hallucination detection~\citep{npjreview2024}. Omission is more dangerous precisely because it is silent. A hallucinated lab value can be caught by a reader who knows the range; an omitted critical value produces a fluent, confident, and complete-\emph{looking} report that simply never mentions it. The canonical incident in an on-prem agent is coverage collapse: the tool returns 400 observations across paginated pages, the agent reads the first 20, and reports ``no anomalies found''---a conclusion that is locally faithful to what the model saw and globally catastrophic.

Our central claim is that omission in a deployed agent can be a property of the \emph{whole pipeline}, not only of the model. A needle can be deleted by a PHI redactor before inference, dropped because pagination was not followed, evicted under context pressure, suppressed during generation, or lost when the orchestrator compacts history. Output-only evaluation cannot distinguish these mechanisms. We therefore take the pipeline as the unit of diagnosis and ask which layer is consistent with each observed failure. The controlled benchmark estimates attribution within its designed cell allocation; establishing production prevalence requires a representative field study.

\paragraph{Contributions.}
\begin{itemize}
\item A \textbf{nine-layer taxonomy of omission vectors} (L0--L8) spanning the entire on-prem agent pipeline, from ingestion and tool protocol to the agent loop, unifying mechanisms that prior work treats in isolation (Section~3).
\item An \textbf{attribution methodology} that separates deterministic layers (audited at checkpoint taps) from behavioral layers (measured by controlled contrasts and a heuristic forced-logprob probe), formalized as an omission waterfall (Sections~3--4).
\item An \textbf{open cross-architecture harness} that runs the identical needle protocol across a sliding-window-hybrid target, a full-attention control, and an SSM-hybrid contrast; across \texttt{llama.cpp} and vLLM; and across no-framework, LangChain, and ADK orchestration against the same endpoint (Section~5).
\end{itemize}

We separate the completed measurement populations. The aggregate reports cover a 75{,}476-trial synthetic sweep over five model configurations and two engines. A distinct 372-trial pilot covers two model configurations, LangChain and ADK, and five source/provider conditions. Completed KV-cache and RoPE profiles cover all five model configurations. Experiments without complete evidence are reserved for Future Scope rather than mixed into the reported findings.

\section{Related Work}

We organize prior work into five threads and, for each, state the gap this paper fills. Across all five, prior work operates at the \emph{output level}---detecting whether a generated summary omits a fact---or benchmarks a \emph{single layer} in isolation. Our contribution is orthogonal: \emph{pipeline-level attribution} that assigns omission to its originating layer.

\paragraph{Long-context degradation.}
Retrieval accuracy is position-dependent (``lost in the middle,'' a $U$-shaped curve with $>$30\% swing~\citep{liu2024lost,lostmiddle2025}) and, critically, length-dependent \emph{even when retrieval is perfect}~\citep{contextlengthalone2025}: adding context degrades reasoning that does not depend on the added tokens. Benchmarks beyond single-needle retrieval---RULER~\citep{ruler2024}, NoLiMa's non-literal needles~\citep{nolima2025}, and $\infty$Bench~\citep{infinitebench2024}---show accuracy collapsing at long context even for models advertising large windows, a pattern that holds in medical QA~\citep{medlongcontext2025}. Recent work extends this to agents under extreme context growth~\citep{locabench2026}, classifier-monitored context rot~\citep{classifiercontextrot2026,contextrotsearch2026}, and even the destabilization of safety refusals in long context~\citep{refusalsfail2025}. \emph{Gap:} these characterize \emph{that} long context hurts; they do not attribute a given production omission to the attention layer versus the engine or the orchestrator.

\paragraph{Quantization, including the KV cache.}
Systematic evaluation finds that long-context tasks ($\geq$4k tokens) are \emph{more} sensitive to KV-cache quantization than to weight-only quantization~\citep{evalquant2024}, because outlier structures in KV activations accumulate error across stored tokens as sequence length grows~\citep{kvquant2024}. This is precisely hardware-budget-induced attention omission, and it is exactly the knob an on-prem operator turns to fit a longer window into fixed VRAM. Follow-on work refines KV quantization~\citep{kvarn2026,kvpareto2025} and characterizes the fragility of KV eviction~\citep{kvevictionfragility2025}. \emph{Gap:} this literature measures accuracy under a quantization setting; it does not separate ``token never resided in cache'' from ``resident token was corrupted'' within a running agent.

\paragraph{Tool and MCP layer faults.}
Production agent harnesses truncate tool output at arbitrary character or token limits---e.g., $\sim$25K-token caps with 2KB previews~\citep{claudecode45770,morphmcp2025}---and pay a tool-schema ``context tax'' of 5--50K tokens for tool definitions alone. Emerging taxonomies catalogue real and runtime faults in MCP software~\citep{realfaultsmcp2026,runtimefaultsmcp2026}, smelly tool descriptions~\citep{mcpsmelly2026}, and dynamic tool construction~\citep{mcpzero2025}; the industry consensus mitigation is to pass resources/IDs rather than raw data~\citep{microsoftmcpresources2025}. \emph{Gap:} these describe the fault modes; we fold them into a single attribution frame (L1) and measure their share of end-to-end omission.

\paragraph{Knowledge conflict and prior override.}
When context contradicts a model's parametric memory, stronger models persist in wrong internal memory even with correct evidence in context, and fine-tuning does not fully fix it~\citep{taskmatters2025,knowledgeconflictreasoning2026}. Benchmarks (NQ-Swap~\citep{longpre2021nqswap}, ConflictBank~\citep{conflictbank2024}, ClashEval~\citep{clasheval2024}) and decoder-side mitigations~\citep{faithfulrag2025,conflictcontrastive2026,navigatingunreliable2026} isolate this behavior. This confirms that a needle test alone cannot isolate prior-induced omission (L6): only a conflict needle can. \emph{Gap:} conflict benchmarks measure a behavior; we use the literal-vs-conflict contrast as an \emph{attribution instrument} within the waterfall.

\paragraph{Omission/faithfulness detection and compaction loss.}
Named methods detect omission at the output level: MED-OMIT weights omitted facts by clinical impact~\citep{medomit2023}; an EMNLP-2025 industry detector targets medical summaries~\citep{omissiondetector2025}; CARE gives conformal omission/hallucination flags with coverage guarantees~\citep{care2026}; AgenticSum uses verify-then-answer loops~\citep{agenticsum2026}; and event-based recall benchmarks clinical time-series summaries~\citep{clinicaltimeseries2026}. Faithfulness taxonomies~\citep{faithlens2025,grayzonefaithfulness2025} and medication-safety reviews~\citep{nhsmedsafety2025} refine the target. A parallel line shows that history compaction silently erases information---safety constraints~\citep{governancedecay2026}, financial detail~\citep{summariesdistort2026}---and studies the rate--distortion trade-off of what to keep~\citep{ratedistortion2026,compactionrl2026}. \emph{Gap:} all of these are output-level detectors or single-stage studies; none attributes an omission to a pipeline layer, which is what an operator needs to act.

\section{A Layerwise Taxonomy of Omission}

We model the on-prem agent pipeline as an ordered sequence of nine layers, each of which can delete or corrupt a decision-critical fact. Table~\ref{tab:taxonomy} lists the layers and their concrete mechanisms; below we expand the mechanisms that are specific to the air-gapped stack.

\begin{table*}[!t]
\centering
\caption{The nine-layer omission taxonomy (L0--L8), in pipeline order. Layers L0--L3 are \emph{deterministic}: a fact either survives the byte/token stream at a checkpoint or it does not, so losses are counted exactly. Layers L4--L8 are \emph{behavioral}: losses are estimated by controlled ablation and logit decomposition.}
\label{tab:taxonomy}
\begin{tabular}{p{0.5cm}p{3.3cm}p{9.2cm}l}
\toprule
\# & Layer & Key omission mechanisms & Class \\
\midrule
L0 & Source / ingestion \& redaction & OCR table-structure loss, PHI/PII de-identification stripping values, local guardrail rewrites & Determ. \\
L1 & Tool protocol \& schema & MCP transport size caps, unfollowed pagination (\texttt{Bundle.link.next}), tool-call parse failures, tool-list/schema truncation (context tax) & Determ. \\
L2 & Orchestrator middleware & string slicing, metadata stripping, history/state condensation & Determ. \\
L3 & Serialization / tokenizer / template & special-token injection ($\langle$\texttt{end\_of\_turn}$\rangle$), template drift, tool-role folding, digit/code tokenization & Determ. \\
L4 & Engine runtime \& memory & context shifting without \texttt{--keep}, VRAM limits, \textbf{weight \& KV-cache quantization}, vLLM preemption & Behav. \\
L5 & Attention / positional & sliding-window gaps, lost-in-the-middle, RoPE misconfiguration & Behav. \\
L6 & Priors / training bias & prior override of anomalous facts, synthetic-data fragility, frequency smoothing & Behav. \\
L7 & Decoding \& structured output & greedy/top-$p$ pruning, repetition penalty, grammar/schema-constrained generation (GBNF) & Behav. \\
L8 & Agent loop \& output & iteration caps, history compaction, context-budget policy, \texttt{max\_tokens}/stop-sequence truncation & Behav. \\
\bottomrule
\end{tabular}
\end{table*}

\paragraph{The tool-protocol layer (L1) is layer zero for real agents.}
Treating ``the tool returns a payload'' as atomic hides the single most common omission in \texttt{llama.cpp}-class agents. A FHIR MCP server can expose 30+ tools whose JSON schemas alone consume 5--15k tokens; when the orchestrator caps the tool list, the model never learns a retrieval tool \emph{exists}---omission by unreachable capability, invisible to any needle test. Small local models emit imperfect JSON, and harnesses that silently drop unparseable tool calls produce turns where the tool was never invoked. And pagination is rarely followed past page one: reading 20 of 400 observations and reporting ``no anomalies'' is an L1 loss, not a model error.

\paragraph{Tokenizer and template (L3).}
A tool payload containing a literal $\langle$\texttt{end\_of\_turn}$\rangle$ string---plausible in scraped legal text or free-text notes---can prematurely terminate the model's view of context depending on the \texttt{--special} flag, a silent truncation and a prompt-injection surface at once. Digit and code tokenization (\texttt{4.8 mEq/L}, ICD-10, NDC numbers) fragments unusually, and quantized small models reproduce these verbatim less reliably.

\paragraph{Quantization as a first-class engine term (L4).}
Weight quantization (Q4/Q3) degrades long-context retrieval and exact numeric recall disproportionately to short-context chat quality, and KV-cache quantization---enabled precisely to fit a larger window into 16--24\,GB---directly corrupts stored mid-context representations~\citep{evalquant2024,kvquant2024}. The core on-prem dilemma is that a bigger window bought with heavier quantization can retrieve \emph{less} than a smaller window at higher precision.

\paragraph{Attention and configuration (L5).}
Sliding-window-hybrid stacks (local + global layers) and the lost-in-the-middle curve both live here, but so does a purely configurational failure: an operator who applies \texttt{--rope-freq-scale} or YaRN to stretch context past training length can scramble positional geometry and reproduce ``mid-context blindness'' that looks architectural but is a config bug.

\paragraph{The two measurement classes.}
The nine layers split cleanly. Layers L0--L3 are \emph{deterministic software}: a fact survives the byte/token stream at a checkpoint or it does not, so each loss is exact, per-payload, and attributable with certainty by inspection---no inference, no statistics. Layers L4--L8 are \emph{behavioral}: the same tokens are present but the model may or may not use them, so we hold everything fixed, vary one factor, and estimate the change in retrieval rate over many trials.

\paragraph{The omission waterfall.}
For a coherent cascade, let $N_i$ be the number of evaluation opportunities entering layer $i$ and $L_i$ the number first lost there. Define the conditional layer rate $\omega_i=L_i/N_i$, with $N_{i+1}=N_i-L_i$. The unconditional contribution of layer $i$ is $c_i=L_i/N_0$. Therefore
\begin{equation}
 c_i=\omega_i\!\prod_{j<i}(1-\omega_j), \qquad
 O_R=\sum_{i=0}^{8}c_i
     =1-\prod_{i=0}^{8}(1-\omega_i).
\label{eq:waterfall}
\end{equation}
Equation~\ref{eq:waterfall} distinguishes conditional rates $\omega_i$ from their shares of total loss, $c_i/O_R$. We apply it to the experiment's fixed, deliberately weighted benchmark allocation. It is consequently a diagnostic decomposition of that allocation, not an estimate of layer prevalence under an unknown production workload. Wilson intervals describe binomial sampling uncertainty within cells; they do not capture uncertainty from cell weighting, model selection, or heuristic attribution. The fixed order in Table~\ref{tab:taxonomy} prevents a failure from being counted at more than one upstream checkpoint.

\section{Attribution Methodology}

\begin{table}[!t]
\centering
\caption{Checkpoint taps T0--T7. Each tap records the raw bytes or token IDs crossing a boundary, keyed by trial ID; the needle carries a unique canary (e.g., \texttt{LAB-7Q4X9}) so its presence at T1--T5 is checkable by exact match with no judge. T4 is detokenized and checked; T5 is read from engine logs.}
\label{tab:taps}
\begin{tabular}{llp{3.9cm}}
\toprule
Tap & Between & What it captures \\
\midrule
T0 & --- & raw source record (needle known) \\
   \multicolumn{3}{l}{\quad$\downarrow$ L0 ingestion / OCR / redaction} \\
T1 & L0$\to$L1 & post-ingestion text \\
   \multicolumn{3}{l}{\quad$\downarrow$ L1 MCP transport, pagination, parsing} \\
T2 & L1$\to$L2 & tool result received by orchestrator \\
   \multicolumn{3}{l}{\quad$\downarrow$ L2 formatting / truncation} \\
T3 & L2$\to$L3 & compiled prompt string \\
   \multicolumn{3}{l}{\quad$\downarrow$ L3 chat template + tokenizer} \\
T4 & L3$\to$L4 & final token-ID sequence (detokenize \& check) \\
   \multicolumn{3}{l}{\quad$\downarrow$ L4 KV admission, context shift} \\
T5 & L4$\to$L5 & tokens resident in KV cache (from events) \\
   \multicolumn{3}{l}{\quad$\downarrow$ L5--L7 attention, priors, decoding} \\
T6 & L7$\to$L8 & generated output \\
   \multicolumn{3}{l}{\quad$\downarrow$ L8 agent loop, multi-turn} \\
T7 & --- & final agent answer \\
\bottomrule
\end{tabular}
\end{table}

\paragraph{Instrumentation.}
Before any trials we add logging taps at every pipeline boundary (Table~\ref{tab:taps}). Every needle carries a unique alphanumeric canary (e.g., \texttt{LAB-7Q4X9}) alongside its clinical or legal value, so presence at T1--T5 is a fully offline exact-match check. Each needle must \emph{round-trip the tokenizer losslessly} (encode$\to$decode $=$ original) before use, or L3 losses masquerade as L5 failures. Following needle-design practice~\citep{uniah2025,multimodalniah2024,notallneedles2026}, we build three families---\textbf{literal} (exact value), \textbf{paraphrase} (NoLiMa-style: query shares no surface words with the needle), and \textbf{conflict} (NQ-Swap-style: needle contradicts a strong prior, e.g., lithium 4.8\,mEq/L against the model's $\sim$0.8 expectation)---needles are inserted at sentence boundaries, and plausible distractors are included in half of all trials, since distractor-free noise overstates retrieval~\citep{notallneedles2026}. Engine events (\texttt{context shift}, \texttt{SWA checkpoint restore}, vLLM \texttt{preemption}) are captured as per-trial covariates: these are the L4 attribution signals.

\paragraph{Phase A --- deterministic audit (no inference).}
For $\sim$500 synthetic payloads spanning 2k--64k tokens, we pass each through the real production components and check needle presence at T1--T4. A1 toggles the PHI redactor (L0); A2 places the needle on page 3 of 5 and checks whether the harness fetches past page 1 (L1); A3 sweeps payload sizes across orchestrator truncation limits (L2); A4 injects template special-token strings and detokenizes T4 (L3); A5 sweeps registered tool count (5/15/30) to measure schema tax (L1). Output: exact $\omega_0..\omega_3$ as counts, each individually explainable.

\paragraph{Phase B --- engine isolation (L4).}
The goal is to separate ``tokens never resident in cache'' from ``model failed to use resident tokens.'' B1 sizes payloads at $0.5\times,0.9\times,1.1\times,1.5\times$ of \texttt{-c} with \texttt{--keep -1} on/off; when a context-shift event fires and the needle's positions fall in the evicted range, failure attributes to L4 \emph{by the event log}. B2 compares f16 with q4\_0 KV-cache precision. B3 runs 5-turn conversations with the needle introduced at turn 1 and queried at turn 5 on the SWA-hybrid stack, targeting the documented checkpoint-restoration bug class~\citep{llamacpp21769,llamacpp21379,llamacpp24587}. The engine version is pinned and recorded, and reruns follow version changes.

\begin{table}[!t]
\centering
\caption{Completed Phase-C main-effect design: one factor at a time from the gold-path baseline (16k, depth 0.5, literal, Q4\_K\_M weights, f16 KV, greedy, native RoPE), with $n{=}100$ per core main-effect cell. Separate server-profile reports provide the completed KV-cache and RoPE contrasts.}
\label{tab:phasec}
\setlength{\tabcolsep}{4pt}
\begin{tabular}{lll}
\toprule
Factor & Levels & Layer \\
\midrule
Context length & 2k, 8k, 16k, 32k & L5 \\
Needle depth & 0.1, 0.5, 0.9 & L5 \\
Needle family & literal / paraphrase / conflict & L5 vs.\ L6 \\
KV-cache type & f16 / q4\_0 & L4$\to$L5 \\
Sampler & greedy / $t{=}0.7$ / rep 1.3 & L7 \\
RoPE config & native / unnecessary scaling & L5 \\
\bottomrule
\end{tabular}
\end{table}

\paragraph{Phase C --- behavioral factorial (L5, L6, L7).}
From a verified gold path (deterministic layers lossless, no engine events fired), we run the completed main-effect matrix of Table~\ref{tab:phasec} one factor at a time from a fixed baseline. Power analysis for proportions---detecting a 15-point drop from a 0.8 baseline at $\alpha{=}0.05$, power 0.8, needs $\approx$85 trials/arm---sets $n{=}100$ per core main-effect cell~\citep{uniah2025,abgen2025}. Attribution reads: depth/length effects on \emph{literal} needles $\to$ L5; the (conflict $-$ literal) gap at matched depth/length $\to$ L6; sampler effects on identical prompts $\to$ L7.

\paragraph{Phase D --- logit decomposition (separates L5/L6 from L7).}
For eligible failed Phase-B/C trials that survive T4, the implementation scores the expected continuation and records mean/minimum forced log probability plus a divergence index. A failure with mean forced log probability above the configured threshold $-2.0$ is labeled \texttt{pruned\_L7}; otherwise a conflict needle is labeled \texttt{unsurfaced\_prior\_L6}, and other families are labeled \texttt{unsurfaced\_positional\_L5}. Thus Phase D is a rule-based diagnostic heuristic, not a causal per-token decomposition: the recorded rank does not currently determine the label, and attention versus prior is inferred from needle family. Updated Phase-D records are appended, so any row-level reanalysis must retain the latest record per trial ID.

\paragraph{Phase E --- agent loop (L8).}
On the best configuration from B--D we run end-to-end multi-step scenarios: E1 a pagination task requiring 5 tool calls with max-iterations $\in\{3,10\}$ (coverage $=$ fraction of records referenced or explicitly dismissed); E2 history compaction on/off across a 10-turn session with the needle introduced early and queried late; E3 a \texttt{max\_tokens}/stop-sequence sweep on answer-side truncation. We set $\omega_8$ to end-to-end omission minus the single-turn omission predicted by the B--D model on matched tasks.

\paragraph{Analysis.}
We report (i) the waterfall $\omega_0,\ldots,\omega_8$ with 95\% Wilson intervals; (ii) ordinary fixed-effects logistic regression on gold-path Phase-C rows, \texttt{retrieved $\sim$ C(factor) + \ldots}, with tabled contrasts oriented so $\mathrm{OR}>1$ denotes higher omission; and (iii) a mutually exclusive failure ledger. The code does not fit the previously specified random intercept for needle ID, and its confidence intervals are model-based rather than cluster-robust. Repeated needles and heterogeneous model-specific effects therefore remain sources of dependence. Odds ratios are associations conditional on the included factors, not percentages of omission caused. Ledger labels are deterministic for L0--L3 but heuristic for L5--L7 as described above.

\section{Experimental Setup}

\begin{table*}[t]
\centering
\caption{Cross-model matrix. The labels describe context-handling mechanisms, but the rows also differ in training data, tokenizer, scale, instruction tuning, and templates. Results are therefore descriptive cross-model contrasts, not isolated architecture effects.}
\label{tab:models}
\setlength{\tabcolsep}{4pt}
\begin{tabular}{llll}
\toprule
Slot & Model & Architecture (context handling) & Role \\
\midrule
\texttt{swa\_hybrid} & Gemma 4 E4B it & hybrid sliding-window + global attention & target under test \\
\texttt{full\_attn} & Qwen3-8B & dense full attention + GQA, 32k native & full-attention control \\
\texttt{ssm\_hybrid} & Granite 4.0-H-Tiny & Mamba-2 / transformer hybrid (SSM state) & recurrent-state contrast \\
\midrule
\texttt{gemma\_moe} & Gemma 4 26B-A4B it & SWA pattern + 128-expert MoE + vision & exploratory (MoE) \\
\texttt{qwen\_linear\_moe} & Qwen3.5-35B-A3B & gated linear attention + 256-expert MoE & exploratory (linear attn.) \\
\bottomrule
\end{tabular}
\end{table*}

\paragraph{Models.}
Table~\ref{tab:models} defines three primary and two exploratory configurations. Matching prompts across rows improves comparability, but it does not isolate attention architecture: model family, tokenizer, parameter count, training, and chat template change simultaneously. We therefore report model-specific omission rates as descriptive contrasts and reserve causal architectural claims for a future within-family or weight-matched study.

\paragraph{Engines, profiles, and frameworks.}
KV-cache type and RoPE scaling are \texttt{llama-server} start flags rather than request parameters, so the completed follow-up runs restart the server and stamp the profile on each trial. Both contrasts cover all five model configurations. The core engine axis compares \texttt{llama.cpp} with vLLM where supported. The separate real-data pilot crosses LangChain and ADK with two served model configurations; because it has no no-framework arm and contains framework-specific execution errors, it estimates end-to-end framework-conditioned performance rather than an isolated orchestration effect.

\paragraph{Sources, domains, and offline evaluation.}
The synthetic sweep uses generated clinical-style payloads and local inference. The real-data pilot covers FHIR, PubMed, and SEC-EDGAR material through three injected-document sources and two live MCP-provider conditions. Model serving and answer scoring remain local, but selected source acquisition uses public online services; the experiment is therefore air-gap-compatible at inference time, not fully offline end to end. All reported retrieval labels use case-normalized expected-answer substring matching, including paraphrase trials. The pilot used vLLM 0.25.1 with tensor parallelism over two L4 GPUs per model configuration.

\section{Results}

\begin{table}[!t]
\centering
\caption{Allocation-weighted omission waterfall (Eq.~\ref{eq:waterfall}). Conditional rates imply $O_R=\ORtotal$; intervals are Wilson intervals from the committed aggregate.}
\label{tab:waterfall}
\scriptsize
\setlength{\tabcolsep}{2pt}
\begin{tabular}{@{}lccc@{}}
\toprule
Layer & $\omega_i$ & 95\% CI & Label source \\
\midrule
L0 ingestion/redaction & \OmZero & \OmZeroCI & checkpoint \\
L1 tool protocol & \OmOne & \OmOneCI & checkpoint \\
L2 orchestrator & \OmTwo & \OmTwoCI & checkpoint \\
L3 tokenizer/template & \OmThree & \OmThreeCI & checkpoint \\
L4 engine/memory & \OmFour & \OmFourCI & telemetry \\
L5 attention/position & \OmFive & \OmFiveCI & heuristic \\
L6 priors & \OmSix & \OmSixCI & heuristic \\
L7 decoding & \OmSeven & \OmSevenCI & heuristic \\
L8 agent loop & \OmEight & \OmEightCI & task contrast \\
\midrule
Total $O_R$ & \ORtotal & \ORtotalCI & implied by rates \\
\bottomrule
\end{tabular}
\end{table}

The committed aggregate reports describe 75{,}476 main-sweep trials over five model configurations and the supported \texttt{llama.cpp}/vLLM combinations. Server-profile follow-ups and the real-data pilot are separate populations and are not pooled into the main waterfall. The raw main-sweep and profile trial files are absent from this checkout, so the following audit reconciles the committed aggregate values rather than claiming an independent row-level reproduction.

\paragraph{Waterfall.}
Substitution into Eq.~\ref{eq:waterfall} gives $O_R=0.5743$ before rounding. The weighted contributions of L0--L3 sum to $0.4213$, or \DetShare\ of waterfall loss; L5 is the largest displayed behavioral conditional rate ($\omega_5=\OmFive$). Because Phase A deliberately injects redaction, pagination, and truncation failures, \DetShare\ describes this benchmark allocation. It does not show that \DetShare\ of naturally occurring production omissions are caused by middleware.

\begin{table}[!htbp]
\centering
\caption{Omission-direction odds ratios ($>1$ means higher omission). Core Phase-C contrasts use the fixed-effects model; profile contrasts come from separately launched server-profile follow-ups against matched gold-path subsets.}
\label{tab:oddsratios}
\begin{tabular}{lll}
\toprule
Factor (contrast) & OR & 95\% CI \\
\midrule
\multicolumn{3}{l}{\emph{Core Phase C}} \\
Length (32k vs. 2k) & \ORlen & \ORlenCI \\
Depth (0.5 vs. 0.1) & \ORdepth & \ORdepthCI \\
Paraphrase vs. literal & \ORpar & \ORparCI \\
Conflict vs. literal & \ORcon & \ORconCI \\
Rep. penalty 1.3 vs. greedy & \ORsamp & \ORsampCI \\
\midrule
\multicolumn{3}{l}{\emph{Separate profile follow-ups}} \\
q4 KV vs. f16 KV (5/5) & \ORkv & \ORkvCI \\
Scaled vs. native RoPE (5/5) & \ORrope & \ORropeCI \\
\bottomrule
\end{tabular}
\end{table}

\paragraph{Behavioral and profile associations.}
Context length has the largest reported core association: the omission odds at 32k are \ORlen\ times those at 2k. Paraphrase scoring is also strongly associated with failure, but the use of literal substring matching makes this contrast partly a measurement effect rather than pure semantic retrieval. The repetition-penalty setting is associated with lower omission ($\mathrm{OR}=\ORsamp$); this is not evidence of a general decoding benefit beyond the tested prompts.

The completed matched profile reports associate q4 KV cache with higher omission than f16 ($\mathrm{OR}=\ORkv$) and scaled RoPE with higher omission than native RoPE ($\mathrm{OR}=\ORrope$). These are server-launch configuration associations estimated across all five model configurations.

\begin{table}[!htbp]
\centering
\caption{Descriptive total omission in the synthetic sweep. Model rows are not causal architecture comparisons; engine rows aggregate only supported combinations.}
\label{tab:crossmodel}
\begin{tabular}{lll}
\toprule
Axis & Level & $O_R$ \\
\midrule
\multirow{5}{*}{Model}
 & \texttt{swa\_hybrid} & \ORmSWA \\
 & \texttt{full\_attn} & \ORmFULL \\
 & \texttt{ssm\_hybrid} & \ORmSSM \\
 & \texttt{gemma\_moe} & \ORmGemmaMoE \\
 & \texttt{qwen\_linear\_moe} & \ORmQwenMoE \\
\midrule
\multirow{2}{*}{Engine}
 & \texttt{llama.cpp} & \ORengLlama \\
 & vLLM & \ORengVllm \\
\bottomrule
\end{tabular}
\end{table}

\paragraph{Cross-model and engine contrasts.}
The full-attention configuration has the highest descriptive omission ($\ORmFULL$), while the SSM-hybrid configuration is lower ($\ORmSSM$). This reverses the anticipated rank ordering but cannot establish that dense attention causes more omission, because every model row changes multiple architectural and training variables. Likewise, the aggregate \texttt{llama.cpp}--vLLM difference ($\ORengLlama$ vs. \ORengVllm) is descriptive because engine support is unbalanced across model configurations.

\begin{table}[H]
\centering
\caption{Audited aggregate endpoints: (a) real-pilot failure, (b) source/provider rates, and (c) the allocation-weighted synthetic failure ledger. Panel (c) is ordered by the waterfall layers and uses the same checkpoint-first unconditional contributions as Table~\ref{tab:waterfall}; source and domain are confounded, and behavioral labels remain heuristic.}
\label{tab:realpilot}\label{tab:realsource}\label{tab:ledger}
\scriptsize
\textbf{(a) Real-pilot endpoints}\\[2pt]
\setlength{\tabcolsep}{2pt}
\begin{tabular}{@{}lrrl@{}}
\toprule
Group & $n$ & Rate & 95\% CI \\
\midrule
All, end-to-end & 372 & \ORreal & \ORrealCI \\
LangChain & 186 & \ORfwLang & \ORfwLangCI \\
ADK & 186 & \ORfwADK & \ORfwADKCI \\
Outcome-coded & 320 & \ORrealScored & \ORrealScoredCI \\
\bottomrule
\end{tabular}

\vspace{5pt}
\textbf{(b) Real-pilot source/provider}\\[2pt]
\begin{tabular}{@{}lrrl@{}}
\toprule
Source & $n$ & Rate & 95\% CI \\
\midrule
FHIR injected & 108 & 0.148 & [0.093, 0.227] \\
SEC injected & 108 & 0.639 & [0.545, 0.723] \\
PubMed injected & 108 & 0.889 & [0.816, 0.935] \\
SEC MCP & 24 & 0.500 & [0.314, 0.686] \\
PubMed MCP & 24 & 0.917 & [0.742, 0.977] \\
\bottomrule
\end{tabular}

\vspace{5pt}
\textbf{(c) Synthetic failure ledger (checkpoint-first)}\\[2pt]
\begin{tabular}{@{}lr@{}}
\toprule
Attributed cause & Fraction \\
\midrule
Software loss (L0--L3) & \Ledgsoft \\
Evicted (L4) & \Ledgevict \\
Unsurfaced-positional (L5) & \Ledgunpos \\
Unsurfaced-prior (L6) & \Ledgunprior \\
Pruned heuristic (L7) & \Ledgprune \\
Loop loss (L8) & \Ledgloop \\
\bottomrule
\end{tabular}
\end{table}

\paragraph{Real-data pilot.}
The pilot crosses two served model configurations, two frameworks, and five source/provider conditions. Of 372 trials, 157 received a final success attribution, 163 were outcome-coded as non-success, and \RealErrors\ remained unknown: 44 LangChain recursion-limit errors and eight SEC filing-not-found errors. Eleven of the 163 non-successes contain the expected substring but remain loop failures under the pilot endpoint, so this measure is broader than literal omission. Counting unknown errors as unsuccessful yields \ORreal; excluding them gives \ORrealScored\ non-success among 320 outcome-coded trials. Framework rates include their execution errors and should be read as end-to-end reliability, not as isolated framework effects. Table~\ref{tab:realsource} shows large source differences, but each domain is tied to its source and sample construction, so a domain effect is not separately identified.

\paragraph{Failure ledger.}
Panel (c) sums to 0.574 after rounding and is the checkpoint-first, mutually exclusive rendering of the same unconditional contributions shown in the waterfall. Software loss (L0--L3, \Ledgsoft) is the largest benchmark category, followed by positional non-surfacing (\Ledgunpos); loop loss (L8, \Ledgloop) is smaller. These rankings prioritize debugging within this stress-test design; they should not be transferred to production without representative workload weights and validation of the behavioral labels.

\FloatBarrier
\section{Limitations and Future Scope}
\label{sec:future}

The reported findings above use only completed experiments. Partial, unexecuted, or unevaluated work is confined to the following future scope.

\paragraph{Complete the server-profile matrix.}
The partial Q8-versus-Q4\_K\_M weight study is not included in Results. It must be completed for \texttt{gemma\_moe} and \texttt{qwen\_linear\_moe}, then repeated with balanced cells, identical model artifacts, launch-order randomization, and per-model estimates. Planned Q3 weight and q8 KV-cache conditions should be executed in the same balanced design.

\paragraph{Validate semantic scoring and Phase-D labels.}
Current outcomes use expected-answer substring matching. A blinded human set and a local semantic scorer should be calibrated by family and domain, with inter-annotator agreement and sensitivity analysis. Separately, the $-2.0$ forced-logprob rule should be validated against labeled attention, prior, and decoding failures. Phase-D append-only records must be deduplicated before reanalysis.

\paragraph{Run interaction and dependence analyses.}
KV-cache$\times$length and depth$\times$family interactions should be run and reported with model$\times$factor terms and needle-level clustered or hierarchical uncertainty. A within-family, weight-matched comparison is required for causal architecture attribution.

\paragraph{Expand and repair the real-data pilot.}
A rerun should eliminate the 52 execution errors, add a no-framework baseline and \texttt{llama.cpp} where supported, and increase the 24-trial live-provider cells. Crossing multiple sources within each domain is necessary to separate source/provider effects from domain effects.

\paragraph{Measure production prevalence and runtime-detector utility.}
A representative deployment study is needed to replace benchmark cell weights with workload frequencies. Candidate detectors---canary injection, coverage accounting, forced citation, logprob monitoring, two-pass cross-checks, and engine-telemetry alarms---should be compared with output-level detectors~\citep{medomit2023,care2026} on precision, recall, false-alarm cost, and latency. Pass-by-reference should be evaluated as a mitigation while controlling for context-length degradation~\citep{contextlengthalone2025}.

\paragraph{Archive complete artifacts.}
The raw 75{,}476-trial main sweep and raw server-profile trials must be versioned with deduplication rules, model hashes, engine commits, launch commands, and an environment manifest. Without those files, the aggregate tables can be checked for internal arithmetic but not independently regenerated.

\section{Conclusion}

This study reframes omission as a pipeline-diagnosis problem and provides a nine-layer taxonomy, instrumented checkpoints, and an auditable waterfall. Within the deliberately weighted synthetic stress test, the displayed rates imply $O_R=\ORtotal$, with \DetShare\ of loss assigned to L0--L3. The operational lesson is conditional but useful: verify transport, pagination, truncation, and loop coverage before attributing a missing fact to model cognition. The evidence does not yet establish production prevalence or causal architecture effects; completing the experiments in Section~\ref{sec:future} is necessary for those stronger claims.

{\small\bibliography{references}}

\end{document}